\documentstyle[12pt]{article}

\begin{document}
\title{\textbf{Magnetic filamentary structures in the spectrum of kinematic dynamos in  plasmas}} \maketitle
{\sl \textbf{L.C. Garcia de Andrade}\newline
Departamento de F\'{\i}sica
Te\'orica-IF\newline
Universidade do Estado do Rio de Janeiro\\[-2mm]
Rua S\~ao Francisco Xavier, 524\\[-2mm]
Cep 20550-003, Maracan\~a, Rio de Janeiro, RJ, Brasil\\[-1.5mm]
Electronic mail address: garcia@dft.if.uerj.br\\[-1.5mm]
\vspace{0.01cm} \newline{\bf Abstract} \paragraph*{}Kinney et al [PPL \textbf{1},(1994)] have investigated plasma filamentary structure dynamics. More recently, Wilkin et al [Phys Rev Lett \textbf{99}:134501,(2007)] have shown that kinetic energy spectrum of magnetic structures in small-scale dynamos, are predominantly filamentary. Kirilov et al [PRE (2009)] have shown that use of the boundary values of the mean-field isotropic helical turbulent ${\alpha}^{2}$-dynamo, could determine the morphology of dynamo instability. In this paper, filamentary Frenet frame in diffusive media, displays the existence of kinematic chaotic dynamo in 3D Euclidean space ${\textbf{E}^{3}}$. In 2D, either no dynamo action spectrum is found, in agreement with Cowling anti-dynamo theorem, or slow dynamos [PPL \textbf{15},(2008)]. Curvature and diffusion effects are encodded in the matrix representation of ${\alpha}^{2}$-dynamo operator. Instead of principal scalar curvatures ${\kappa}_{1}$ and ${\kappa}_{2}$ of the surface of structures, only one scalar curvature ${\kappa}$ is needed to determine dynamos spectra. Filament thickness, increases with scalar curvature, as happens in solar physics.
\newpage
\section{Introduction}
Filamentary structures in MHD plasma dynamics have been investigated by Kinney et al \cite{1}, who have used the concept of Elss\"{a}sser variables. Their approach was also based on the vortex dynamics structures. In their paper, however, no dynamo action in plasma was investigated. More recently Wilkin et al \cite{2} have investigated the existence of dynamo action in turbulent filamentary structures, showing that the small-scale dynamos \cite{3}, can actually produce these filamentary structures. They also showed that, at least in kinematic stage of dynamos, filamentary profiles are preffered rather than surface structures like ribbons. In their study they made use of the Reynolds magnetic number $R_{m}=\frac{vl}{\epsilon}$, where v and l are respectively, the typical velocities and scales involved in the plasma, while ${\epsilon}$ is the magnetic diffusity. Typically in galactic and astrophysical plasmas, l is very big and highly conductive media $({\epsilon}\approx{0})$, leads to high magnetic Reynolds numbers. Critical $R_{m}$ are within the range of $50-500$. This critical magnetic numbers are the minimum numbers able to support dynamo action. On the other hand, previous investigations by Guenther et al \cite{4} on ${\alpha}^{2}$-dynamo operator spectrum indicates that a simple eigenvalue analysis similar to the spectrum analysis of isotropic spherically symmetric Dirac and Schroedinger operators, and the consequent investigation of the eigenvalue growth rate of the magnetic fields, leads one naturally to the behaviour of dynamo action. Actually in their work, they also showed that the simple analysis of polynomial in three-dimensional space can be useful in this investigation. In this paper, following the approach of Guenther et al, one is able to show that a yet simpler eigenvalue analysis can be obtained for ${\alpha}^{2}$-dynamo when one write the induction equation in the Frenet local reference frame, which follows isolated filaments. Thus by assuming the filamentary structure of Wilkin et al, one is led to a dynamo action in three-dimensions. In two-dimensional regime, either the dynamo action cannot be supported or at the best, slow dynamos in plasmas \cite{5} is obtained. Considerable effort to provide simple Riemannian geometrical dynamo models \cite{6,7}, have been made recently, since the first toy model of a chaotic dynamo on a torus surface, given by Arnold et al \cite{8}. One of the drawbacks of simple models is that in general they do not need to fast dynamo action as appears in solar plasmas and galactic dynamos. Nevertheless maximum possible simplicity, might be an important caractheristic to be addressed. Another interesting aspect of the application of differential Riemannian geometry to dynamos is that Anosov two dimensional constant Riemannian curvature spaces has been demonstrated by Chicone et al \cite{7} to be good candidates of kinematic fast dynamos in highly conductive ideal plasmas \cite{9}. Actually it is shown that the simple Frenet frame used in this paper, the eigenvalue spectra is very similar to the one obtained by Chicone and Latushkin, though main differences between their work and ours, consists in the facts that, they use a dynamo equation which is not a ${\alpha}^{2}$-dynamo; secondly their analysis used a differential forms approach to obtain the dynamo spectrum on surfaces and not filamented dynamos in Frenet frame obtained here. From the mathematical viewpoint, the distinction comes from the fact that here, one uses the differential geometry of curves in three-dimensional Euclidean spaces, rather than the Riemannian geometry of the two-dimensional surfaces in $\textbf{E}^{3}$. The paper is organised as follows: In section II the Frenet frame formalism is applied to two dimensional ${\alpha}^{2}$-dynamos. In section III the same spectral analysis is applied to chaotic, dynamo operators is invoked in three dimensions. Section IV presents future prospects and conclusions.
\newpage
\section{Filamentary structures in slow 2D-${\alpha}^{2}$-dynamos}
This section presents the main ideas on the spectrum of the induction equation, and subsequent investigation on the possible existence of the ${\alpha}^{2}$-dynamo in $2D$. The magnetic lines along the filaments are computed in the Frenet frame $(\textbf{t},\textbf{n},\textbf{b})$, where the tangent vector $\textbf{t}$, is along these lines, while the respectively, normal and binormal vectors $\textbf{n}$ and $\textbf{b}$ belong to an orthogonal plane to the magnetic filament. This frame vectors obey the following evolution equations
\begin{equation}
\frac{d\textbf{t}}{ds}={\kappa}(s)\textbf{n}
\label{1}
\end{equation}
\begin{equation}
\frac{d\textbf{n}}{ds}=-{\kappa}(s)\textbf{t}+{\tau}\textbf{b}
\label{2}
\end{equation}
\begin{equation}
\frac{d\textbf{b}}{ds}=-{\tau}(s)\textbf{n}
\label{3}
\end{equation}
Here ${\kappa}$ and ${\tau}$ are Frenet curvature and torsion scalars. The magnetic self-induction equation is written as
\begin{equation}
\frac{{\partial}\textbf{B}}{{\partial}t}={\nabla}{\times}({\alpha}\textbf{B})+
{\epsilon}{\Delta}\textbf{B}\label{4}
\end{equation}
where ${\alpha}=<\textbf{v}.{\nabla}{\times}\textbf{v}>$ represents the ${\alpha}$ helicity of the flow given by $\textbf{v}=v_{0}\textbf{t}$. Note that though one considers here that the modulus of the flow $v_{0}$ is considered as constant the flow is not necessarily laminar due to the dynamical unsteady nature of the frame vector $\textbf{t}$. This equation shall be expanded below, along the Frenet frame as
\begin{equation}
\textbf{B}(s,t)=B_{s}(s,t)\textbf{t}(t,s)+B_{n}\textbf{n}+B_{b}\textbf{b}\label{5}
\end{equation}
Since in this section, one shall be considering the 2D case one shall assume from the beginning that the binormal component of the magnetic field $B_{b}$ shall vanish. Another simplification one shall addopt here is that the normal and binormal magnetic perturbations $B_{n}$ and $B_{b}$ shall be considered as constants. Some technical observations are in order now. The first is that this kind of Frenet frame used here, are called the isotropic Frenet frame, which considerers that even the frame is unsteady as here, in the sense that they depend upon time, the base vectors only depends upon the toroidal coordinate-s, and not other coordinates. Otherwise the Frenet frame is called anisotropic. The anisotropic may be more akin and suitable to turbulent phenomena but is much more involved, and shall be left to a next paper. Some simple use of the anisotropic Frenet frame can be found in simple Arnold like dynamos in reference \cite{10}. Let us start the MHD equations by the solenoidal divergence-free vector field by
\begin{equation}
{\nabla}.\textbf{B}=0 \label{6}
\end{equation}
as
\begin{equation}
{\partial}_{s}B_{s}-{\kappa}_{0}B_{n}=0 \label{7}
\end{equation}
Note that ${\kappa}_{0}={\tau}_{0}$ here represents the constant torsion ${\tau}_{0}$ equals the curvature ${\kappa}_{0}$ of the helical filamentary turbulence. By computing the relations
\begin{equation}
{\Delta}\textbf{t}=-{{\kappa}_{0}}^{2}\textbf{t}\label{8}
\end{equation}
\begin{equation}
{\Delta}\textbf{n}=-{{\kappa}_{0}}^{2}{\textbf{n}}\label{9}
\end{equation}
Since the 2D ${\alpha}^{2}$-dynamos is embedded in $\textbf{E}^{3}$, decomposition of the induction equation (\ref{4}), might be done along the three base vectors of the Frenet frame. By considering the expression
\begin{equation}
{\partial}_{t}\textbf{B}=({\lambda}B_{s}-{\kappa}_{0}B_{n})\textbf{t}+({\lambda}B_{n}+
{\kappa}_{0}(B_{s}-B_{n}))\textbf{n}-{\kappa}_{0}B_{n}\textbf{b}
\label{10}
\end{equation}
where the time evolution of $\textbf{B}$ has the Lyapunov chaotic behaviour is $|\textbf{B}|=e^{{\lambda}t}$. Besides from the corresponding curl of the helicity term, one obtains the respective eigenvalue equations along $\textbf{t}$, $\textbf{n}$ and $\textbf{b}$ directions, are
\begin{equation}
[{\lambda}+{{\kappa}_{0}}^{2}{\epsilon}]B_{s}-{\kappa}_{0}B_{n}=0\label{11}
\end{equation}
\begin{equation}
[{\lambda}+{{\kappa}_{0}}^{2}{\epsilon}]B_{n}=0\label{12}
\end{equation}
\begin{equation}
[{\alpha}{\kappa}_{0}B_{s}+({\partial}_{s}{\alpha})B_{n}]=0\label{13}
\end{equation}
Since one is in 2D one shall consider the last expression as a constraint on the ${\alpha}$ helicity. As far as the degeneracy of the scalar components of the magnetic is concerned one is faced with two posibilities:i) The first is when they are degenerate, or $B_{s}=B_{n}$; in this case equation (\ref{13}) reduces to
\begin{equation}
({\alpha}{\kappa}_{0}+{\partial}_{s}{\alpha})B_{s}=0\label{14}
\end{equation}
Since $B_{s}$ does not vanish by assumption, a simple solution of this constraint to the ${\alpha}$ helicity is
\begin{equation}
{\alpha}={\alpha}_{0}e^{-{\kappa}_{0}s}\label{15}
\end{equation}
which shows that curvature effects tend to kill helicity effects, and besides ${\alpha}$ is positive as in the Parker's cyclonic dynamo effect. Nevertheless, this last expression also shows that helicity is unstable and vanishes as the one goes through the filament \cite{10} as $s\rightarrow{0}$. Unfortunatly, here there is no dynamo at all since from the remaining equations (\ref{12}) and (\ref{13}) yields through the eigenvalue spectrum equation
\begin{equation}
det[{\lambda}\textbf{I}-{D}_{\epsilon}]=0\label{16}
\end{equation}
Here $\textbf{I}$ represents the unit matrix
\vspace{1mm}
\begin{equation}$$\displaylines{\pmatrix{1&{0}\cr{0}&
1\cr}\cr}$$\label{17}
\end{equation}
where two-dimensional possible, ${\alpha}$-dynamo operator matrix $D_{\epsilon}$ can be written as
\begin{equation}
\vspace{1mm} $$\displaylines{\pmatrix{{\epsilon}{{\kappa}_{0}}^{2}&{-{\kappa}_{0}}\cr{0}&
{-{\epsilon}{{\kappa}_{0}}^{2}}\cr}\cr}$$
\label{18}
\end{equation}
Now by considering the matrix of eigenvalues in the form
\begin{equation}
\vspace{1mm} $$\displaylines{\pmatrix{{\lambda}+{\epsilon}{{\kappa}_{0}}^{2}&{-{\kappa}_{0}}\cr{0}&{{\lambda}-{\epsilon}{{\kappa}_{0}}^{2}}\cr}\cr}$$
\label{19}
\end{equation}
one obtains a second-order algebraic equation, which in the degeneracy case, where the discriminant ${\Delta}$ of the algebraic equations vanishes yields
\begin{equation}
{\lambda}_{\pm}=-2{\epsilon}{{\kappa}_{0}}^{2}\le{0}\label{20}
\end{equation}
The vanishing of the growth rate ${\lambda}$ of the magnetic field, leads us to two possibilities: the first is that curvature vanishes, and this is a trivial impossible case in turbulent filaments; the second is the limit ${\epsilon}\rightarrow{0}$ which is essentially the slow dynamo limit.
\section{Filamentary structures in 3D kinematic dynamo spectra}
Let us now consider the magnetic kinematic dynamo, which considers the regular induction equation in 3D with non-zero plasma resistivity ${\epsilon}$
\begin{equation}
\frac{{\partial}\textbf{B}}{{\partial}t}={\nabla}{\times}[\textbf{v}{\times}\textbf{B}]+
{\epsilon}{\Delta}\textbf{B}\label{21}
\end{equation}
where ${\Delta}:={\nabla}^{2}$ is the Laplacian operator. Here we also assume that the same decomposition as above is done, with the difference that now the binormal component of the magnetic field $B_{b}$ does not vanish. Besides here the magnetic helicity does not appear in the induction equation. In this case the divergence free law or the absence of magnetic monopole remains the same. By considering the rescaling $v_{0}:=1$, the three scalar induction equation obtained from the decomposition of the vector induction equation along the filaments is, along $\textbf{t}$, $\textbf{n}$ and $\textbf{b}$ directions, are
\begin{equation}
[{\lambda}-{{\kappa}_{0}}^{2}{\epsilon}]B_{s}-{\kappa}_{0}B_{n}+{\epsilon}{{\kappa}_{0}}^{2}B_{b}=0\label{22}
\end{equation}
\begin{equation}
{\kappa}_{0}B_{s}+[{\lambda}+2{{\kappa}_{0}}^{2}{\epsilon}]B_{n}-2{\kappa}_{0}B_{b}=0\label{23}
\end{equation}
\begin{equation}
-{\alpha}{{\kappa}_{0}}^{2}B_{s}-2{\kappa}_{0}B_{n}+({\lambda}+{\epsilon}{{\kappa}_{0}}^{2})
=0\label{24}
\end{equation}
In this 3D dynamo, the eigenvalue spectrum equation
\begin{equation}
det[{\lambda}\textbf{I}-{D}_{\epsilon}]=0\label{25}
\end{equation}
where now the dynamo operator matrix $D_{\epsilon}$, can be written as
\begin{equation}
$$\displaylines{\pmatrix{{-{\epsilon}{{\kappa}_{0}}^{2}}&{-{\kappa}_{0}}&
{{\epsilon}{{\kappa}_{0}}^{2}}\cr{{\kappa}_0}&{2{\epsilon}{{\kappa}_{0}}^{2}}&
{-2{\kappa}_{0}}\cr{-{\epsilon}{{\kappa}_{0}}^{2}}
&{-2{\kappa}_{0}}&{{\epsilon}{{\kappa}_{0}}^{2}}\cr}\cr}$$
\label{26}
\end{equation}
As in section II, the matrix of eigenvalues leads to the third-order algebraic equation
\begin{equation}
{\lambda}^{3}-{\epsilon}{{\kappa}_{0}}^{2}{\lambda}^{2}-{{\kappa}_{0}}^{2}
[1+\frac{{\epsilon}}{2}{{\kappa}_{0}}^{2}]{\lambda}+2[1-\frac{1}{2}
[1-4{{\kappa}_{0}}{\epsilon}]]
{{\kappa}_{0}}^{2}=0
\label{27}
\end{equation}
Though in general, complete third-order algebraic equations and higher, are very complicated one shall address here a special case of physical interest to dynamo theory. In this case the equation is reduced to a second-order equation. In this case, one shall consider that the growth rate of the magnetic field ${\lambda}$ is very small on a kind of slow dynamo. Thus, the third-order term in ${\lambda}$, could be truncated, or neglected  and the polynomial equation (\ref{27}) would be reduced to
\begin{equation}
-{\epsilon}{{\kappa}_{0}}^{2}{\lambda}^{2}-{{\kappa}_{0}}^{2}
[1+\frac{{\epsilon}}{2}{{\kappa}_{0}}^{2}]{\lambda}+2[1-\frac{1}{2}
[1-4{{\kappa}_{0}}{\epsilon}]]
{{\kappa}_{0}}^{2}=0
\label{28}
\end{equation}
A simple particular solution of this equation can be obtained as
\begin{equation}
{\lambda}=2^{\frac{1}{3}}[1-\frac{1}{2}(1-4{{\kappa}_{0}}{\epsilon})]{{\kappa}_{0}}^{\frac{2}{3}}\label{29}
\end{equation}
which in the limit of ideal plasmas where resistivity ${\epsilon}$ vanishes, one obtains
\begin{equation}
lim_{{\epsilon}\rightarrow{0}}{\cal{R}}{\lambda}=[{\frac{{{\kappa}_{0}}}{2}}]^{\frac{2}{3}}
\label{30}
\end{equation}
Here, ${\cal{R}}{\lambda}$, is the real part of ${\lambda}$ which in general has complex roots, indicating that the dynamos oscillates. Thus since this limit is positive, and does not vanish (slow dynamo), a fast dynamo solution is obtained from this dynamo spectrum. In the above computations the incompressible flows
\begin{equation}
{\nabla}.\textbf{v}=0
\label{31}
\end{equation}
\newpage
are compatible with the above definition of the flow. Just for comparison, one shall reproduce here the Riemannian three dimensional spectrum obtained by Chicone et al as
\begin{equation}
$$\displaylines{\pmatrix{{-{\epsilon}}&{0}&
{0}\cr{0}&{-{\epsilon}}&
{-{\kappa}_{0}}\cr{{\epsilon}{{\kappa}_{0}}^{2}}
&{1-{\kappa}_{0}{\epsilon}}&{-{\epsilon}{{\kappa}_{0}}^{2}}\cr}\cr}$$
\label{32}
\end{equation}
One may note that these matrices are very similar in carachter and dependence on the magnetic diffusivity constant and constant curvature. Chicone et al did not considered in detail this three dimensional dynamo and no mention has been done on turbulent ${\alpha}^{2}$-dynamo.
\section{Conclusions}
 By making use of mathematical tools from operator spectral theory, an investigation of the spectra of two kind of dynamos is performed. The first is the ${\alpha}^{2}$-dynamo, so useful in turbulence and geodynamos. The second is the chaotic dynamo so useful in MHD dynamo plasma theory. A new fast dynamo solution comes out from this spectrum investigation, where the eigenvalue spectrum is obtained from a particular solution of the third-order algebraic equation. A more complete panorama of the present solution can be obtained by performing the graphic between the growth rate of the magnetic field and the Reynolds magnetic number $Rm$. Since the Rm is the inverse of the parameter of diffusion ${\epsilon}$ certainly the fast dynamo presented here implies a high Rm, which by the relation obtained by Wilkin et al \cite{2}, for the filaments thickness $l_{\epsilon}=l_{0}{Rm}^{-\frac{1}{2}}$, one may conclude that the fast dynamo obtained in the last section is actually a thin filament dynamo with thickness which depend upon curvature as $l_{\epsilon}=2{{\kappa}_{0}}^{\frac{1}{2}}l_{0}$, which shows that the filament Since curvature is related to the folding, one may say that a faster dynamo can be obtained in three-dimensions with the enhancement of folding, what does not happen in two-dimensions.\section{Acknowledgements}
 Several discussions with Dmitry Sokolov are highly appreciated. I also thank Andrew Soward for kindly sending me a reprint of his work on slow dynamos. Financial  supports from UERJ and CNPq are gratefully acknowledged.
 
  \end{document}